# Effects of Interface Disorder on Valley Splitting in SiGe/Si/SiGe Quantum Wells

Zhengping Jiang,[1] Neerav Kharche,[2] Timothy Boykin,[3] Gerhard Klimeck[1]

[1] *Network for Computational Nanotechnology, Purdue University, West Lafayette, IN 47907, USA*

[2] *Computational Center for Nanotechnology Innovations, Rensselaer Polytechnic Institute, Troy, NY 12180, USA*

[3] *Department of Electrical and Computer Engineering, University of Alabama in Huntsville, Huntsville, AL 35899, USA*

A sharp potential barrier at the Si/SiGe interface introduces valley splitting (VS), which lifts the 2-fold valley degeneracy in strained SiGe/Si/SiGe quantum wells (QWs). This work examines in detail the effects of Si/SiGe interface disorder on the VS in an atomistic tight binding approach based on statistical sampling. VS is analyzed as a function of electric field, QW thickness, and simulation domain size. Strong electric fields push the electron wavefunctions into the SiGe buffer and introduce significant VS variations from device to device. A Gedankenexperiment with ordered alloys sheds light on the importance of different bonding configurations on VS. We conclude that a single SiGe band offset and effective mass cannot comprehend the complex Si/SiGe interface interactions that dominate VS.

Quantum computing (QC) is being actively pursued because of its potential functionalities beyond today's classical computers.[1-4] Quantum computers differ from classical computers in their unique states of quantum bits (qubits). The qubit initialization and the requirement of long decoherence time have been the main challenges for the material and architecture selection for a quantum computer.[5] Silicon has been considered as one of the most promising materials to build qubits[2,6] due to the availability of a spinless isotope $^{28}$Si and its small spin-orbit coupling, which ultimately should result in long spin decoherence time. In addition the hope is high to utilize the expertise of the entire Si-based semiconductor industry to ultimately deliver high levels of integration with well-developed technology.

One of the most promising QC architectures, which makes use of the electron spin as a qubit is based on (001) oriented SiGe/Si/SiGe heterostructures.[6-11] Such designs rely on exact control of qubit states by electronic gates, hence a non-degenerate ground state is preferred.[3,12] However, Si has six equivalent conduction minima, which brings additional complexity to qubit gate operations. In a (001) SiGe/Si/SiGe heterostructure, biaxial expansion of Si on the SiGe virtual substrate lifts the 6-fold valley degeneracy to lower two-fold and higher four-fold degeneracy. The remaining two-fold degeneracy is still a potential source of decoherence.[2,6,12]

In the vicinity of a sharp potential barrier, the two degenerate valley states at opposite ends of the Brillouin zone will be coupled by valley-orbit coupling and form two states slightly separated in energy. The energy difference is defined as VS. This splitting, however, is very sensitive to the details of interface such as barrier height, strain, miscuts, etc.[13-18] For successful realization of QC devices it is crucial to understand the dependence of VS on the interface details.

In the calculation of VS, the conduction band offset at the Si/SiGe interface is generally considered to be the most important parameter. To reduce the computational burden and to obtain an approximation, the atomistic scale disorder in SiGe is usually either ignored or included by fitting parameters.[13,14,17] This work investigates the effects of disorder quantitatively at the SiGe/Si interface with considerations of the varying atomistic details of a Si interface with an atomistically disordered SiGe barrier material. Up to $5\times10^5$ atoms are included to address strain and electronic properties in real disordered system. The 20-band $sp^3d^5s^*$ nearest neighbor tight-binding (TB) model (no spin-orbit) is utilized to represent atomistically resolved device samples.

The paper is organized as follows. First, the tight-binding model is validated against the experimentally measured conduction and valence band offsets of the bulk SiGe alloy, where a good quantitative match is obtained. Then the dependence of VS in SiGe/Si/SiGe QWs on the Si well thickness and electric field is analyzed. The calculated VS follows the experimentally observed dependence on the QW thickness. To pin down the importance of including the atomic scale disorder in the predictive VS calculations, SiGe barriers with the same Ge composition but different distribution of Ge atoms are simulated. The atomistic details of the Si/SiGe interface are found to play a crucial role in determining VS.

The potential energy barrier at the SiGe/Si interface, which affects the VS in Si QW, is very sensitive to the local electronic structure and the strain distribution at the interface. Here, the strain relaxation is carried out using the valence-force-field (VFF) model with modified Keating potential. The subsequent electronic structure calculation is performed by the $sp^3d^5s^*$ nearest neighbor TB model, which includes strain. The atomistic nature of both the VFF and TB methods enables the accurate modeling of the local variations in strain and electronic structure on an atomistic scale. In all the VS calculations, the strain relaxation domain includes 25 nm thick SiGe layers on both sides of the Si QW while in the electronic structure calculation this thickness is reduced to 4 nm beyond which the wavefunction leakage is negligible. To account for the variation due to the disorder in SiGe barrier, 20 different samples with random atom configurations are averaged to obtain the VS in Si QW. The calculations are performed using NEMO-3D.[19, 20]

*Model Validation in Bulk SiGe:* The VFF and TB models used here have been proved previously to accurately reproduce the band gaps, effective masses, and their dependence on strain for pure Si, pure Ge, and an ordered Si/Ge superlattice.[21-23] To validate these models for the SiGe alloy, where Si and Ge atoms are randomly distributed, in this work the calculated valence and conduction band edges are compared with their experimentally measured values in FIG. 1(a). The band edges are extracted by zone unfolding[22, 24, 25] the bandstructures of SiGe supercells of size $4\times4\times40$ nm$^3$. Our model and experimental data are in close agreement all SiGe alloy compositions, including the *X* and *L* valley crossover.

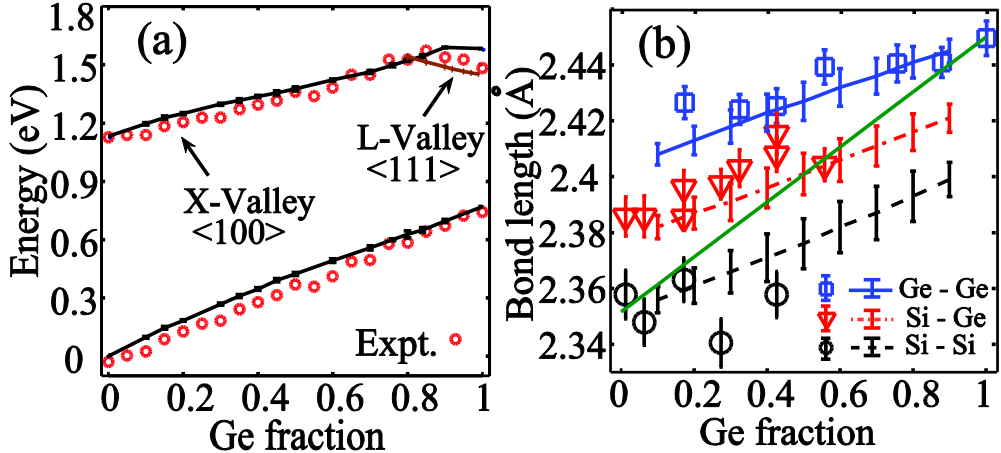

FIG. 1 Validation of TB and strain VFF model by comparison with experimental data. (a) Band edges of bulk SiGe in X and L valleys extracted by unfolding dispersion. (b) Statistics of bond length for SiGe bulk from experiment [[26, 27]] (symbols) and VFF model (lines with error bars) showing tri-model bond length distribution. VCA bond length is plotted in straight line.

The simplest approximation assumes that all atoms separate from each other by an average bond length that corresponds to a particular alloy concentration, resulting in a regularly ordered zincblende crystal. Within framework of TB, the resulting structure is free of strain and each site is occupied by identical atom with TB parameters that are weighted average of Si and Ge bulk values. However, Si-Si, Ge-Ge, and Si-Ge interactions have different strengths and they move the respective bonding partners off the perfect zincblende lattice and result in three distinct bond distributions. This is observed experimentally[26, 27] and faithfully reproduced with our VFF strain model in FIG. 1(b). The lattice constant calculated using the virtual crystal approximation (VCA) is shown for comparison. The VCA can reproduce bulk properties like bandedges,[28] which makes it suitable for transport simulations where three-dimensional carrier behavior dominates. However, the deviation from the ideal lattice sites implies that at a SiGe/Si interface there are also positional disorders which will perturb the electron wavefunction interaction at that interface and therefore influence the VS.

In a recent experimental measurement,[7] the VS in SiGe/Si/SiGe heterostructure QWs in the presence of an electric field was observed to increase with decreasing Si well thickness; Si wells of 4nm(1MV/m), 5.3nm(1.3MV/m), 10nm(2MV/m) and 20nm(2.3MV/m) are measured. Through our calculations, FIG. 2(a) shows the dependence of VS on the QW thickness at low electric fields: 1MV/m (cross) and 2MV/m (solid dot). We increase well thickness by step of one unit cell (4 atom layers) and only plot QWs with thickness close to experimental structures. The calculated VS follows similar trend with the QW thickness as the experimental data without adjustment of any material parameters. The calculated VS is higher than the experiments because the miscut at Si/SiGe interfaces, which are known to suppress VS,[8, 16] are not included in these simulations. Recent experimental efforts[10, 29] focus on the use of flat substrates without miscuts. Here we focus on "flat" quantum wells since the primary goal of this work is study the effects of SiGe alloy disorder.

The statistical VS distributions show that for QWs under 10nm, standard deviations are much larger due to the effects of alloy disorder. It has been shown that in the case of an ordered, abrupt potential barrier, the VS fluctuates with well thickness on atomic layer scale.[15, 30] We argue here that due to the presence of disorder, the wavefunction inhomogeneously penetrates into the barrier and hence the effective well thickness is different among samples. This causes the statistical distribution of VS values.

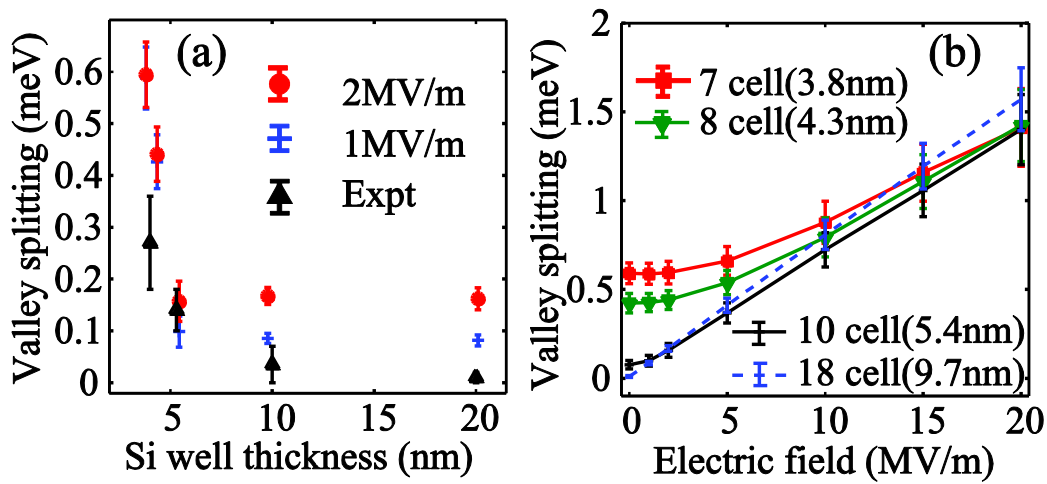

FIG. 2 Well thickness and electric field dependence of VS. (a) VS for different well thicknesses at 1MV/m and 2MV/m. VS from experimental data is shown by triangular markers. VS increases with decreasing well

thickness. (b) Electric field increases VS for all well widths. Disorder has more effects in high electric fields.

In real devices, the electric field due to modulation doping or gate bias could be as large as 10MV/m.[6] FIG. 2(b) depicts the trend of the VS as a function of electric field up to 20MV/m, well thicknesses are labeled by unit cell numbers. Compared with calculations in previous work, VS in the 7 unit cell (~3.8nm) thick, zero field QW is about one third of the VS in an ordered, abrupt barrier.[6, 17] This shows disorder is more than a small correction, but one of the primary effects. As the field increases, VS values also increase, similar to the ordered, abrupt case. The dependence on QW thickness is washed out at higher electric fields because at high fields, wavefunctions pile up at only one interface and are effectively only confined in a triangular well that is independent of the total well length.[15, 30] Interestingly the statistical model shows a more broadened VS distribution with increased electric field providing further indication that the wavefunctions that are pushed deeper into the disordered interface with increased fields experience more disorder effects and associated magnitude variation in the VS.

Sizes of realistic Si quantum dots vary with design and might range from active confinement areas of 10×10 nm$^2$ to 40×40 nm$^2$ or even larger.[31] The natural question arises how large a quantum dot would have to be for disorder induced deviations to become negligible. The results depicted in Figures 1 and 2 focused on relatively small devices with a cross section area of 10×10 nm$^2$ in the X-Y plane over which periodic boundary conditions are enforced. FIG. 3 explores the effects of the simulation cross section size on valley splitting. The fact that the lines in FIG. 3(a) are not flat is indicative that small systems clearly suffer from large sample to sample variations such that the average is not even reaching the large system limit. FIG. 3(b) indicates the decrease of the deviations from large system limit as a function of lateral system size. Large applied electric fields show significantly larger deviations supporting our argument that the wavefunctions explore more spatially varying environment in the SiGe buffer.[18]

To gain a better understanding of the effects of the disordered SiGe we continue[18] to conduct a set of Gedankenexperiments with particular, well defined Si-Ge atom patterns for a 25% Ge concentration. FIG. 4 depicts the prototypical 8-atom unit cell, its atom numbering, and two different samples of two-Ge-atom placements. FIG. 4(b, c, d) show cells in which the Ge atoms occur in positions {4,6}, {5,6} and {2,6}. By using the three unit cell types, effects of Ge-Ge coupling and surface roughness will be evaluated. Firstly, there are no Ge-Ge bonds in the {4,6}, and {5,6} cells and hence barriers built by these unit cells are free of nearest neighbor Ge-Ge bonds. A different confinement is expected compared with {2,6} cell. Secondly, the {4,6} barrier is effectively a superlattice of one Ge and three Si atomic layers, whereas {2,6} and {5,6} barriers are superlattices composed of two $Si_{0.5}Ge_{0.5}$ layers and two Si layers. The {4,6} barrier provides a smoother interface than {2,6} and {5,6}. We repeat the simulation with different well thicknesses in electric fields up to 10MV/m for the fictitious SiGe barriers.

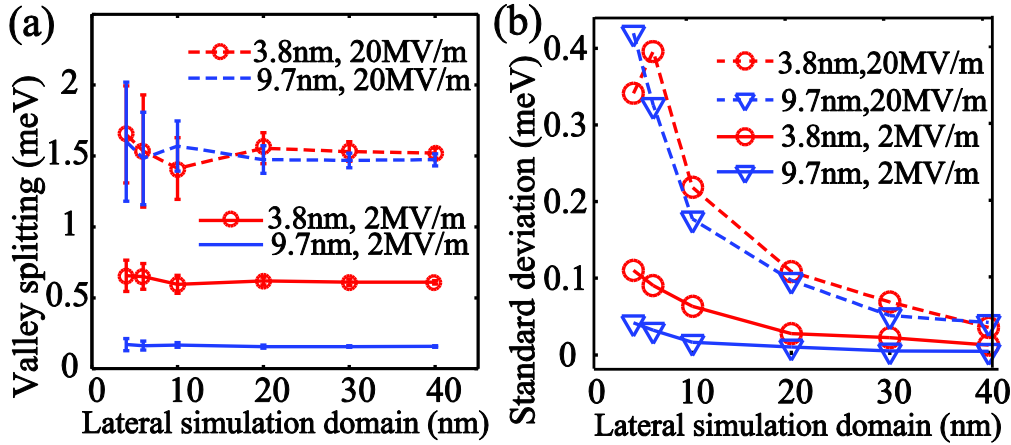

FIG. 3 VS as a function of the non-confining lateral simulation domain size for different quantum well thicknesses and applied electric fields. (b) VS standard deviations as a simulation domain size.

The following systematic VS calculations on Si QWs with ordered SiGe barriers illustrate the effects of having all 3 types of bonds, *viz.*, Si-Si, Si-Ge, and Ge-Ge, on the VS behavior. Firstly FIG. 5(a) compares VS for the {4,6} (solid line with symbols) and {5,6} (dashed line with symbols) barriers for different QW thicknesses against the random barrier results for 3.8nm and 9.7nm wells (shown without error bars for clarity). The VS in the ordered {4,6} and {5,6} barriers is larger than the random barrier case. Then, the VS for {2,6} barriers is plotted in FIG. 5(b) and even though it has the same Ge percentage as {4,6} and {5,6} it contains one nearest-neighbor Ge-Ge bond in each unit cell. With the additional nearest-neighbor Ge-Ge bond, the overall VS values are suppressed and approach values of the fully random case. Finally comparing the {4,6} and {5,6} barriers, the {5,6} barrier shows a slightly lower valley splitting than {4,6} barrier, which is due to a wider and more rough surface, in agreement with effective mass results. This effect is not as large as that due to additional Ge-Ge coupling, though. The similarity of the VS in the {4,6} and {5,6} barriers shows that the presence of Ge-Ge bonds is important: The {4,6} barrier is smooth while the {5,6} barrier, like the {2,6} barrier, is rough.

By applying periodic boundary conditions to cells in FIG. 4(b, c, d), the energy dispersions of the fictitious SiGe barriers are calculated in the confinement direction. We do not show the dispersion here, but conduction band edges extracted from dispersion show that the {4,6} and {5,6} cell produces a similar barrier height as the fully random barrier while the barrier height obtained from {2,6} cell is roughly 100meV lower compared to fully random barrier. Although the {4,6} barrier has similar band offsets as the random barrier, the VS in this case is very different. On the contrary, the {2,6} barrier, which has a different band offset has similar VS as the random barrier. This shows that VS is very sensitive to interface conditions. *Thus, only taking into account the band offset is not sufficient for predictive VS calculations.* Since molecular beam epitaxy (MBE) permits growth of atomic layers of different species,[32] our study provides possible guidelines for future experiments.

In conclusion, in this work we study effects of alloy disorder on VS and highlight importance of interface configurations in determining VS. Our computational models are calibrated by matching bulk material properties and reproducing experimentally observed trends. Effects of disorder are reflected in standard deviation of simulation data. VS is shown to increase with decreasing well thickness. Further study on bond types of SiGe shows that including all bond types is important in understanding VS, matching band offsets is not sufficient to get reasonable VS values. This work explores the underlying physics of Si/SiGe interface from understanding of atom couplings. Continuous models used in other papers[13, 33] and the atomistic model in this work both show a smoother interface can produce more favorable VS for QC from different perspectives.

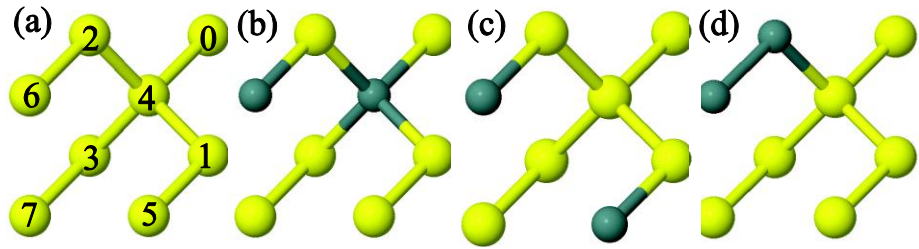

FIG. 4 Examples of 8 atom SiGe unit cells with 25% Ge used to build ordered SiGe barrier. (a) Atoms are numbered from 0 to 7. (b) Unit cell by replacing 4$^{th}$ and 6$^{th}$ Si atom with Ge. Interface with no Ge-Ge bond. (c) Unit cell by replacing 5$^{th}$ and 6$^{th}$ Si atom with Ge. Interface with no Ge-Ge bond. (d) Unit cell with Ge on atom 2 and 6. All three bond types Si-Si, Si-Ge and Ge-Ge are present in this unit cell.

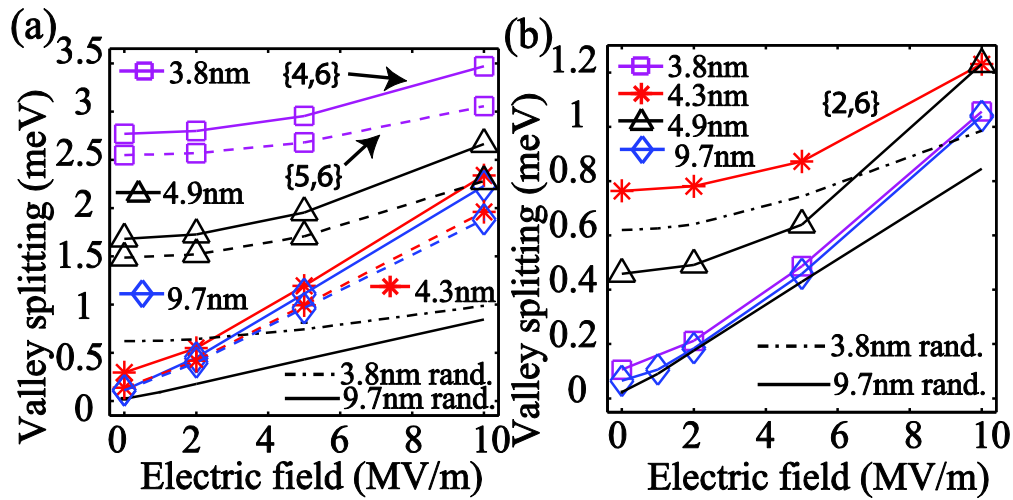

FIG. 5 Effect of bond types in SiGe. (a) VS for SiGe with regular pattern {4,6}(solid lines) and {5,6}(dashed lines). Valley splitting is overestimated without Ge-Ge bond. (b) VS for SiGe with regular pattern {2,6}. {2,6} barriers mimic fully random barriers. Presence of Ge-Ge bonds brings VS values closer to fully random barrier case.

We thank Dr. Michael Povolotsky, Prof. Alejandro Strachan , Dr. Rajib Rahman and Prof. Jim Harris for many valuable discussions. This work is supported by LPS/NSA through ARO award numbers W911-NF-08-1-0482, W911NF-08-1-0527. Computational resources on nanoHUB.org funded by NSF grant EEC-0228390 were used extensively.